\documentclass{PoS}
\usepackage{multirow}
\usepackage{aas_macros}
\usepackage{amsmath, amssymb}

\title{Measurement of the EBL through a combined likelihood analysis of gamma-ray observations of blazars with the MAGIC telescopes}

\ShortTitle{Measurement of the EBL with MAGIC}

\author{\speaker{A. Moralejo Olaizola}$^{1}$,  A. Dom{\'i}nguez$^{2}$, V. Fallah Ramazani$^{3}$, T. Hassan$^{1}$, D. Mazin$^{4,5}$, M. Nievas Rosillo$^{2}$, E. Prandini$^{6}$, J. Sitarek$^{7}$, G. Vanzo$^{8}$, M. V\'azquez Acosta$^{8}$ for the MAGIC collaboration. \thanks{We would like to thank the IAC for the excellent working conditions at the ORM in La Palma. We acknowledge the financial support of the German BMBF, DFG and MPG, the Italian INFN and INAF, the Swiss National Fund SNF, the European ERDF, the Spanish MINECO, the Japanese JSPS and MEXT, the Croatian CSF, and the Polish MNiSzW.}\\
$^{1}$Institut de F{\'i}sica d'Altes Energies (IFAE), Barcelona, Spain\\
$^{2}$Universidad Complutense de Madrid, Grupo de Altas Energ\'{i}as (GAE)\\
$^{3}$Tuorla Observatory, University of Turku and Astronomy Division, University of Oulu, Finland\\
$^{4}$Institute for Cosmic Ray Research (ICRR), University of Tokyo, Japan.\\
$^{5}$Max-Planck-Institut f\"ur Physik, D-80805 M\"unchen, Germany\\
$^{6}$Universit\`a di Padova and INFN, Padova, Italy\\
$^{7}$University of \L\'od\'z, Poland\\
$^{8}$Instituto de Astrof\'isica de Canarias and Universidad de La Laguna, Tenerife, Spain\\
E-mail: \email{moralejo@ifae.es}}

\abstract{The extragalactic background light (EBL) is the radiation accumulated through the history of the Universe in the wavelength range from the ultraviolet to the far infrared. Local foregrounds make the direct measurement of the diffuse EBL notoriously difficult, while robust lower limits have been obtained by adding up the contributions of all the discrete sources resolved in deep infrared and optical galaxy observations. Gamma-ray astronomy has emerged in the past few years as a powerful tool for the study of the EBL: very-high-energy (VHE) photons traversing cosmological distances can interact with EBL photons to produce e$^+$e$^-$ pairs, resulting in an energy-dependent depletion of the gamma-ray flux of distant sources that can be used to set constraints on the EBL density. The study of the EBL is one of the key scientific programs currently carried out by the MAGIC collaboration. We present here the results of the analysis of 32 VHE spectra of 12 blazars in the redshift range 0.03 - 0.94, obtained with over 300 hours of observations with the MAGIC telescopes between 2010 and 2016. A combined likelihood maximization approach is used to evaluate the density and spectrum of the EBL most consistent with the MAGIC observations. The results are compatible with state-of-the-art EBL models, and constrain the EBL density to be within $\simeq 20\%$ the nominal value in such models. The study reveals no anomalies in gamma-ray propagation in the large optical depth regime - contrary to some claims based on meta-analyses of published VHE spectra.}

\FullConference{35th International Cosmic Ray Conference --- ICRC2017\\
		10--20 July, 2017\\
		Bexco, Busan, Korea}

\begin{document}
\section{Introduction}
The Extragalactic Background Light (EBL) is a cosmic diffuse radiation field that encloses essential information about galaxy evolution and cosmology. It is mainly composed by ultraviolet, optical, and near-infrared light emitted by stars and its re-emission to longer wavelengths by interstellar dust, which produces its characteristic double peak spectral energy distribution. This radiation is accumulated over the cosmic history and redshifted by the expansion of the Universe (see \cite{dwek13} for a review, and references therein). Other contributions to the EBL may exist such as those coming from the accretion on super-massive black holes, light from the first stars, or even more exotic sources such as products of the decay of relic dark matter particles. The actual contribution of these components to the total background is poorly known.
\begin{table}[b]
\begin{center}
\begin{tabular}{|l|c|c|c|c|}
\hline
Source  & type & redshift & period & observation\\
 & & & & time (h) \\
\hline
\hline
Markarian 421 (15 spectra) & HBL & 0.030 & 20130410 - 19, 20140426 & 43.8 \\
1ES 1959+650 & HBL & 0.048 & 20151106 - 18 & 4.8 \\
OT 546 (1ES 1727+502) & HBL & 0.055 & 20151012 - 20151102 & 6.4 \\
BL Lacertae & IBL & 0.069 & 20150615 & 1.0 \\
1ES 0229+200 & HBL & 0.14 & 2012 - 2015 & 105.2 \\
1ES 1011+496 & HBL & 0.212 & 20140206 - 20140307 & 11.8 \\
PKS 1510-089 (2 spectra) & FSRQ & 0.361 & 20150518-19, 20160531 & 5.0 \\
PKS 1222+216 & FSRQ & 0.432 & 20100618 & 0.5 \\
PG 1553+113 (5 spectra) & HBL & 0.43 - 0.58 & 2012 - 2016 & 66.4 \\
PKS 1424+240 (2 spectra) & HBL & 0.604 & 2014 - 2015 & 49.1 \\
PKS 1441+25 & FSRQ & 0.939 & 20150418 - 20150423 & 20.1 \\
QSO B0218+35 & FSRQ & 0.944 & 20140725 - 20140726 & 2.1 \\
\hline
\end{tabular}
\end{center}
\vspace{-0.3cm}
\caption{\it List of the 32 MAGIC spectra used in the determination of the EBL density. \label{EBLsampleTable}}
\end{table}
The direct photometric detection of the EBL is challenging because of strong foreground, mainly zodiacal light but also the brightness of our own Galaxy. Therefore, attempts of direct detection are subject to large uncertainties and biases. Other detection analyses focus on measuring the background anisotropies, which still provides inconclusive results (e.g.~\cite{matsumoto05,bernstein07}). These techniques cannot provide any information about the fundamental EBL evolution.

An alternative methodology to estimate the EBL is based on counting photons in different photometric bands using deep-galaxy-surveys data (e.g.~\cite{madau00}). This procedure results in EBL estimates that can be considered robust lower limits. However, cosmic variance may contribute to systematic uncertainties using this technique. Also light from the outer regions of normal galaxies or stars in faint undetected galaxies can be missed.

There are other efforts centered on building empirical models using different complementary methodologies. Following the classification by \cite{dominguez11a} (see references therein), these models are divided in four different classes: (1) Forward evolution models that use semi-analytical models of galaxy formation, (2) Backward evolution models based on local or low redshift galaxy data, which are extrapolated to higher redshifts making some assumptions on the galaxy evolution, (3) Inferred evolution from the cosmic star formation history of the Universe and (4) Observed evolution based on galaxy data over a broad range of redshift. Basically, these models converge to spectral intensities that are close or even match those derived from galaxy counts, at least, in the shorter wavelength peak. Uncertainties are larger in the far-IR peak since most of these models do not include long wavelength data.

Another technique that has become rather successful in constraining the EBL is based on the observation of extragalactic sources of gamma rays. This approach relies on the fact that photons with energies larger than about 10 GeV travelling through cosmological distances suffer attenuation by pair-production interactions with the EBL. By making more or less sophisticated assumptions about the intrinsic/unattenuated spectra, and studying the features of the observed spectra, it is possible to derive information on the EBL and, very importantly, on its evolution. Early attempts provided upper limits to the EBL density (e.g.~\cite{aharonian06}). Yet, more recently, thanks to the availability of more and better gamma-ray data, the {\it detection} of the imprint of the EBL on gamma-ray spectra has been claimed by different groups \cite{ackermann12,abramowski13,dominguez13a,biteau15,1es1011magic}. These EBL detections constrain the background intensities to be near those given by galaxy counts and models (within a factor of about two). However, they are in strong tension with those intensities reported by early direct detection analysis such as the one presented by \cite{matsumoto05} and \cite{bernstein07}, yet still compatible, or slightly in tension, with more recent estimates such as those by \cite{matsuoka11}, \cite{matsuura17}, and \cite{mattila17}.
\par
MAGIC \cite{magicupgrade1, magicupgrade2} is a system of two Imaging Atmospheric Cherenkov Telescopes (IACTs) for very-high energy (VHE) gamma-ray astronomy in the range above $E_\gamma \gtrsim$ 50 GeV. In the past few years, MAGIC and other IACTs have detected VHE emission from sources up to $z \simeq 1$ \cite{pks1441magic,pks1441veritas,b0218}, expanding significantly the range of distances 
 available for gamma-ray attenuation measurements from the ground.

\section{Data sample and analysis method}
The data set presented here consists of 32 VHE spectra from 12 different sources, in the redshift range from z=0.03 to 0.944, obtained with the MAGIC telescopes between 2010 and 2016 (see table \ref{EBLsampleTable}). All the data are taken with MAGIC in stereoscopic mode. The criterion to split the data set of one source into more than one spectrum is based on source spectral shape variability in VHE gamma rays. After quality cuts, the live time of the data set is 316.1 h. We also performed for each data set an analysis of contemporaneous Pass 8 Fermi-LAT public data, from which an estimate of the source flux and photon index at the pivot energy of the Fermi spectrum were obtained. The analysis was performed using the Fermi ScienceTools v10r0p5 \cite{fermianalysis} supplied by the Fermi Science Support Center (FSSC 20150518A) through the {\it Enrico} package \cite{enrico}.
\begin{figure}[htbp]
   \centering
   \includegraphics[width=1.0\linewidth]{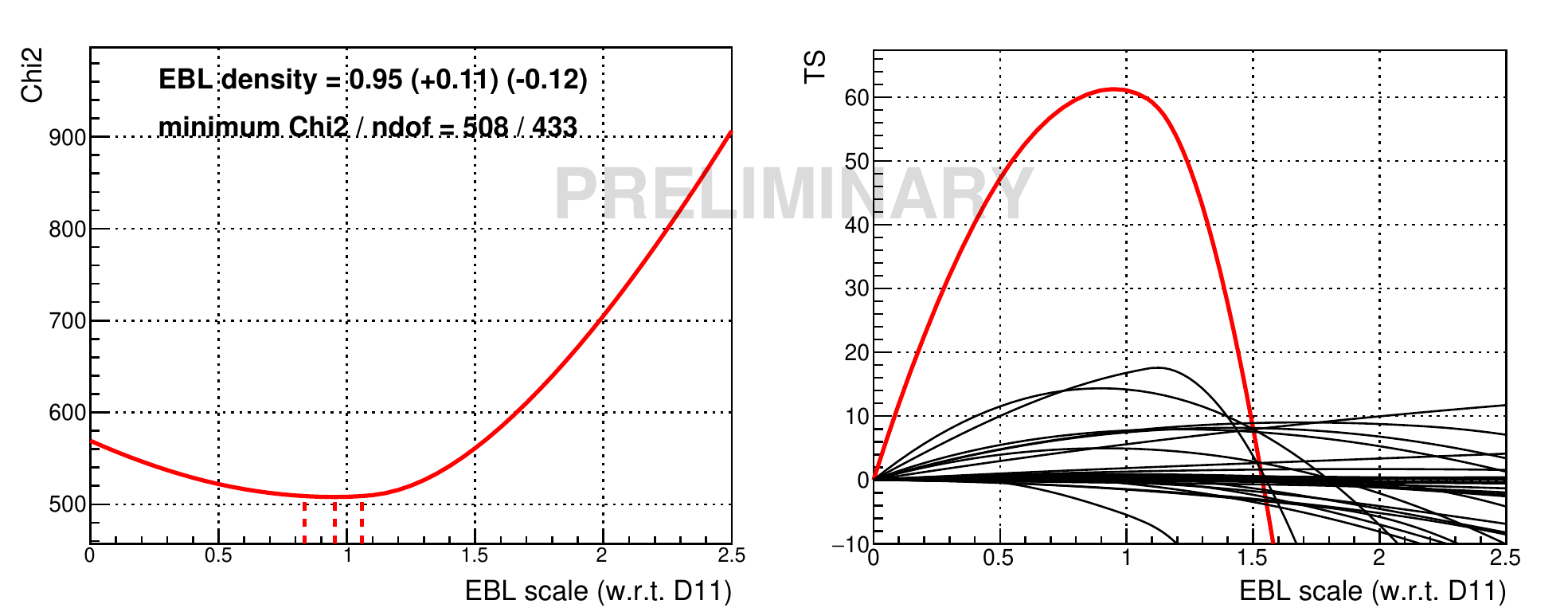}
   \caption{\it Total $\chi^2$ from the maximum likelihood fit (left) and total (red) and spectrum-wise (black) Test Statistics TS $= \chi^2(0) - \chi^2$ (right)  vs. EBL scale relative to the Dom\'inguez model. \label{Result_D11_noPWL_sel2}}
\end{figure}
\par
We follow a procedure similar to the one described in \cite{abramowski13}. The absorption of the EBL is described as $e^{-\alpha \tau(E,z)}$ where $\tau(E,z)$ is the optical depth predicted by a given EBL template model, which depends on the energy E of the gamma rays and the redshift z of the source, and $\alpha$ is a scale factor. The spectrum after EBL absorption is $(dF/dE)_{intrinsic} \; e^{-\alpha \tau(E,z)}$, which is then folded with the MAGIC instrument response function (IRF) obtained from Monte Carlo (MC) simulations, to obtain the expected gamma-ray rates vs. estimated energy $E_{est}$. A binned (in $E_{est}$) Poissonian likelihood is built as the product of the Poisson probabilities of the observed number of events in the sky region around the source, and in three nearby control regions containing only background events. The Poisson parameters of the background, and the statistical uncertainties in the MAGIC IRF (from limited MC statistics) are treated as nuisance parameters as described in \cite{rolke}. The adopted Poissonian maximum likelihood approach works also in the few-events-per-bin regime, hence the full range of estimated energy with non-null bin contents (after cuts) is used, regardless of the significance of the excess. This avoids biases in favour of positive fluctuations in the high end of the MAGIC spectra. 
\par
The likelihood $L$ is then maximized, i.e. $-2 \log L$ minimized, using the MIGRAD algorithm implemented in the ROOT framework\cite{minuit2, root}, with all the parameters describing the intrinsic spectra as free parameters. The redshift of PG 1553+113 (which is uncertain) is also treated as a nuisance parameter in the process, with flat distribution between its lower and its upper limit (as reported in table \ref{EBLsampleTable}). If the maximum value of $L$ in the explored space of parameters is $L_{max}$, and the unconstrained maximum likelihood is $L^*$, then the quantity $-2 \log L_{max}/L^*$ is, in the asymptotic limit, a $\chi^2$ with the number of degrees of freedom of the problem ($N_{Ebins} - N_{parameters}$) \cite{cowan}, which allows to calculate the fit P-value. With this approach we perform a scan of the EBL scale factor, and obtain in each step the best-fit $\chi^2$ (see fig. \ref{Result_D11_noPWL_sel2}). The scale $\alpha_0$ for which the $\chi^2$ reaches its minimum is the value which best fits the data, and the condition $\Delta \chi^2 = 1$ relative to the minimum indicates the $\pm 1 \sigma$ statistical uncertainty range (dashed lines in the figure).
\par
For the combined MAGIC+Fermi analysis we introduce in the $\chi^2$ to be minimized, besides the Poissonian terms, two additional terms for every spectrum, of the form $[(\Gamma - \Gamma_{Fermi}) / \Delta \Gamma_{Fermi}]^2$ and $[(F - F_{Fermi}) / \Delta F_{Fermi}]^2$. Here $F_{Fermi} \pm \Delta F_{Fermi}$ and $\Gamma_{Fermi} \pm \Delta \Gamma_{Fermi}$ are the flux and photon index estimated at the pivot energy of the Fermi-LAT spectrum, whereas $F$ and $\Gamma$ are the flux and photon index obtained from the fitted spectral function. In this way the Fermi data act as an "anchor" which helps constraining the intrinsic HE spectra of the observed sources.
\par
The models for the intrinsic spectra are chosen among concave (in log F vs. log E representation), smooth functions of 3 or 4 parameters: power-law with exponential or sub/super-exponential cut-off (EPWL, SEPWL), log-parabola (LP) and log-parabola with exponential cut-off (ELP). For each spectrum we first adopt the model which achieves the best fit (in terms of P-value) anywhere in the 0-2.5 range of EBL scale factors. A preliminary maximum likelihood estimator of the EBL scale ($\alpha_0 \pm \Delta \alpha_0$) is obtained, and the model selection is revised by repeating the  procedure in the $\alpha_0 \pm 2 \Delta \alpha_0$ range. Whenever two models have the same P-value, the one showing a flatter $\chi^2$ profile around the best-fit EBL (i.e. the one most degenerate with the effect of the EBL) is chosen (as in \cite{1es1011magic}). The pure power-law model (PWL) will only be used (whenever favoured over more complex functions by its P-value) to estimate the systematic uncertainty of the upper EBL bound - such model will naturally bias the result towards high EBL values, by {\it attributing} to the EBL absorption all of the {\it observed} spectral curvature (part of which may be intrinsic).
\begin{figure}[htbp]
   \centering
   \includegraphics[width=1\linewidth]{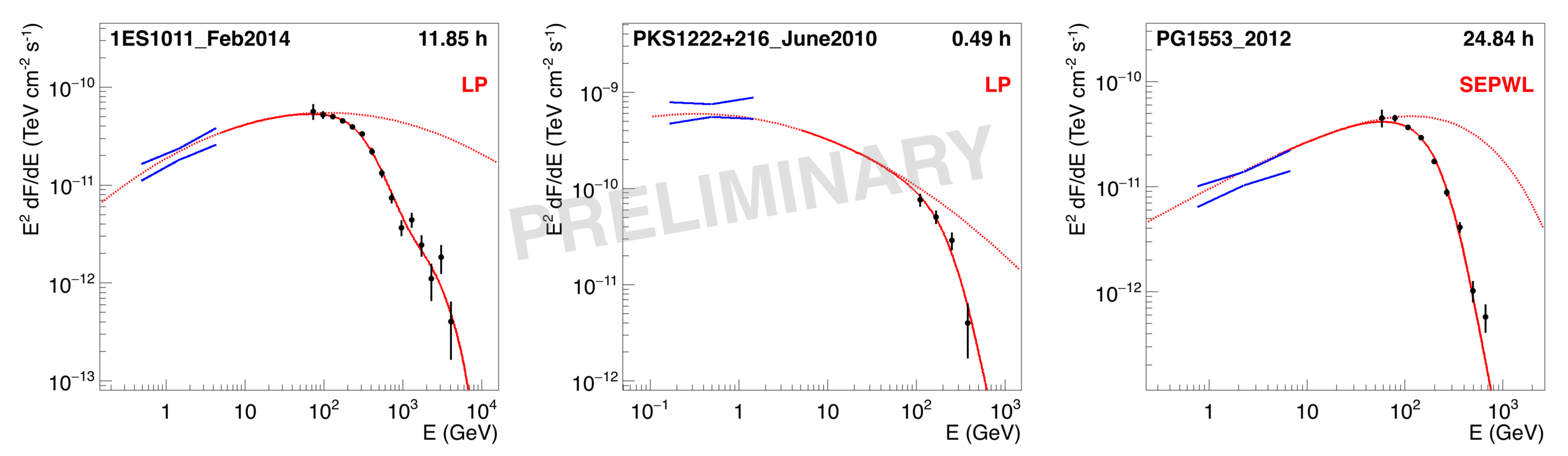}
   \caption{\it Best-fit SEDs for three of the spectra, using the joint Fermi-MAGIC analysis and the D11 EBL template. The blue lines are the bow-ties from the Fermi-LAT analysis. The dotted red lines are the best-fit intrinsic spectra. Solid lines are the corresponding absorbed spectra. The labels in red indicate the spectral model used in each case (see text). \label{ThreeSEDs}}
\end{figure}
\section{Results}
We applied the outlined method to three different EBL model templates: the 2008 model of Franceschini et al \cite{franceschini08} (F08);  the 2011 model of Dom\' inguez et al \cite{dominguez11a} (D11),  and the 2012 fiducial model of Gilmore et al \cite{gilmore12} (G12). The results are summarized in table \ref{resultstable}, both for the MAGIC-only analysis and for the joint MAGIC+Fermi analysis. For  the MAGIC-only analysis best-fit EBL scale factors are all within 1 $\sigma_{stat}$ of 1.0, meaning the data are compatible with the nominal optical depth  (hence EBL density) in the models. The $\chi^2$ profile for the D11 template is shown in fig. \ref{Result_D11_noPWL_sel2}.
\par
\begin{table}[htp]
\begin{center}
\begin{tabular}{|c|c|c||c|}
\cline{2-4}
\multicolumn{1}{c|}{} &   \multicolumn{2}{|c||}{MAGIC-only analysis} & MAGIC with Fermi \\
\hline
EBL & \multirow{2}{*}{Best-fit scale} & allowing PWL & \multirow{2}{*}{Best-fit scale} \\
template &   & as intrinsic spectral model & \\
\hline
\multirow{2}{*}{F08}     & 1.00 $(+0.11, -0.12 )_{stat}$   &  1.13 $(+0.06)_{stat}$  &  0.94 $(+0.07, -0.06 )_{stat}$ \\
                         &  P = $8.8\times10^{-3}$          &  P = $1.9 \times10^{-2}$ &  P = $6.4 \times10^{-5}$\\
\hline
\multirow{2}{*}{D11}     & 0.95 $(+0.11, - 0.12)_{stat}$   &   1.08 $(+0.06)_{stat}$    &   0.91 $(+0.07, -0.06 )_{stat}$ \\
                         & P = $7.5 \times10^{-3}$            &    P = $1.5 \times10^{-2}$ & P = $3.7 \times10^{-5}$  \\
\hline
\multirow{2}{*}{G12}     & 0.98 $(+0.11, -0.12 )_{stat}$    & 1.12 $(+0.07)_{stat}$  &  0.95 $(+0.07, -0.08 )_{stat}$\\
                         & P = $8.1 \times10^{-3}$            & P = $ 1.6 \times10^{-2}$ & P = $8.0 \times10^{-5}$ \\
\hline
\end{tabular}
\caption{\it EBL density constraints using MAGIC and MAGIC+Fermi spectra (preliminary). The analysis which includes PWL as intrinsic spectral model (central column) is used only to estimate systematic uncertainties on the upper bound.\label{resultstable}}
\end{center}
\vspace{-0.5cm}
\end{table}
The results do not favour any of the three models - not surprisingly, since they are are quite similar in the range of wavelengths and redshifts to which the MAGIC data sample is sensitive. When the Fermi-LAT {\it bow-ties} (see fig. \ref{ThreeSEDs}) are used to constrain the intrinsic spectra, the relative statistical uncertainties of the results drop from $\simeq 12\%$ to $\simeq 7\%$, whereas the best-fit values are slightly lower in all three cases: still within 1  $\sigma_{stat}$ of 1.0 for F08 and G12, and marginally off (by $1.3 \sigma_{stat}$) for D11. The most constraining sources are Mrk 421 (15 highly-significant spectra), 1ES 1011+496 (the only spectrum which shows a clear inflection point around 1 TeV), and PG 1553+113 (five different spectra which contribute significantly to the upper constraint, despite the uncertainty in the source redshift). The P-values are low, which probably indicates that the true intrinsic spectra are more complex than assumed (especially over the $\simeq$ 3 orders of magnitude spanned by the MAGIC and Fermi observations). The choice of intrinsic spectral models is indeed one the of the main sources of systematic uncertainties in this type of analysis, the other being the potential departures of the performance of the instrument (including atmospheric effects) with respect to the Monte Carlo simulations from which its response is estimated. The central column of table \ref{resultstable} shows the MAGIC-only results when PWL is included among the pool of possible intrinsic spectral models (and chosen, because of its best P-value, for 12 of the 32 spectra). The best-fit EBL scale is systematically larger for all three models in this case, by an amount comparable to the uncertainty of the original calculation. Since the PWL is the "limit'' case of a concave spectrum, we consider these values just as estimators of the systematic uncertainty on our upper EBL constraint, rather than trustable measurements of the EBL density. 
\par
To estimate the effect of the systematic uncertainty in atmospheric transmission and telescope efficiency we modified the calibration constants used to convert the pixel-wise digitized signals into photoelectrons by $\pm 15\%$ for all spectra at once. Note that spectrum- or night-wise systematic errors in the calibration already contribute to the statistical uncertainties of our measurement - what we are testing now is the effect of an {\it average} miscalibration of the MC and the whole data sample. By analyzing these modified data sets (and also accepting the PWL among the spectral models) we observe a maximum deviation of $+0.19$ in the best-fit EBL scale for the MAGIC-only analysis. On the lower end of the EBL constraint, the uncertainty is determined by the {\it inability} of the selected spectral models to {\it absorb} the EBL imprint on the spectrum as we reduce the scale factor. If we use more complex spectral models (more degenerate with the EBL) the fitted EBL scale gets lower, and the uncertainty on the low-side gets larger. As an extreme case, we repeat the model selection by taking for each spectrum the model which yields the highest P-value in the no-EBL hypothesis, i.e. the model which best fits the {\it observed} spectrum. This approach biases the result towards lower values, e.g. we get $0.83 (+0.14, -0.15)$ for D11. From these considerations we estimate the systematic uncertainty of the EBL scale factors to be $(+0.19, -0.15)$ for the MAGIC-only analysis. Note that in an earlier version of this study \cite{magicgamma2016} the low-end systematics were grossly overestimated due to an error in the production of the calibration-modified samples. For the MAGIC+Fermi analysis the same procedure results in larger systematics, $(+0.28, -0.26)$, perhaps because the modification of the MAGIC calibration, with the Fermi bow-tie acting as an {\it anchor}, introduces a too large distortion of the spectra - under the assumption that the nominal calibration of MAGIC is closer to being correct than either the + or $-15\%$ variants. For the case of the MAGIC+Fermi analysis there is an additional (so far not evaluated) systematic uncertainty resulting from the lack of strict simultaneity of the observations combined with the possible variability of the sources.
\par
\begin{figure}[htbp]
   \centering
   \includegraphics[width=1.\linewidth]{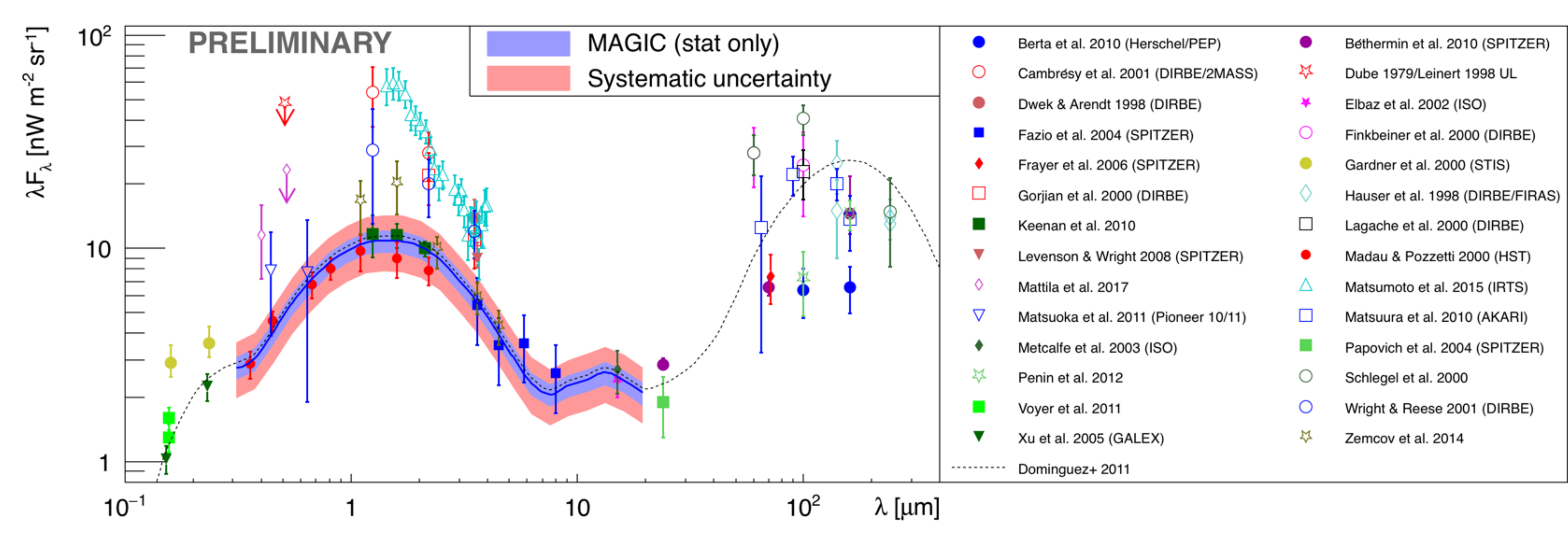}
   \caption{\it Spectral energy distribution of the EBL at z=0 according to the D11 model (dashed line), and scaled by the best-fit factor resulting from the combined analysis of 32 VHE MAGIC spectra of blazars. A selection of direct EBL measurements is shown for comparison. \label{MAGIConly_vs_direct}}
\end{figure}
\begin{figure}[htbp]
   \centering
   \includegraphics[width=1.0\linewidth]{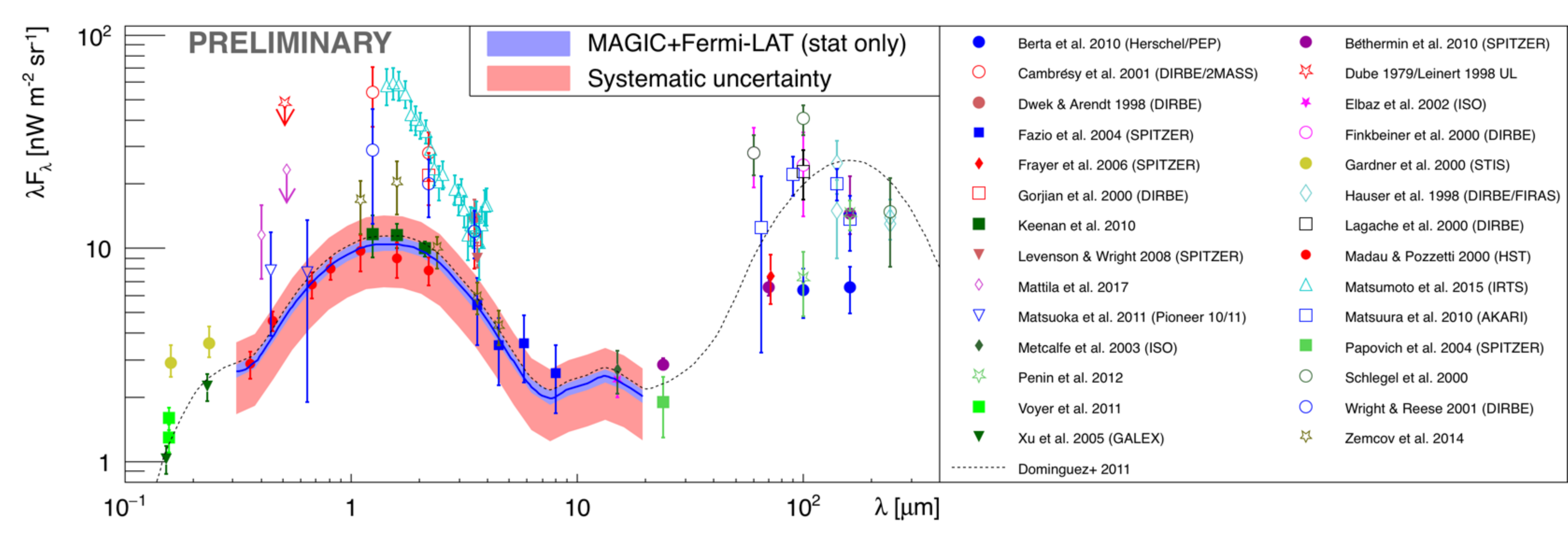}
   \caption{\it Best-fit EBL scaling relative to the D11 model, for the MAGIC + Fermi-LAT analysis of 32 blazar spectra.\label{MAGICandFermivs_direct}}
\end{figure}

The results for the D11 EBL template, with statistical and systematic uncertainties, are shown in figures \ref{MAGIConly_vs_direct} and \ref{MAGICandFermivs_direct}, compared to a number of direct EBL measurements. The high-side of the uncertainty bands is just $\simeq 20\%$ above the lower limits set by galaxy counts (shown as filled symbols in the figures), leaving little room for additional EBL components.
Finally, we have also investigated the behaviour of the spectral fit residuals in the VHE range as a function of the optical depth $\tau$ , in order to test possible anomalies like those claimed in e.g. \cite{horns12} - namely, that observations of distant VHE sources hinted at a reduced opacity of the Universe at high (model-predicted) values of $\tau$. However, no such trend is found in our data set.
%
%
%
%


\begin{thebibliography}{99}
\setlength{\parskip}{0pt}
\setlength{\itemsep}{0pt plus 0.3ex}
\begin{footnotesize}
\bibitem{dwek13} E. Dwek, F. Krennrich, {\it The extragalactic background light and the gamma-ray opacity of the universe}, Astroparticle Physics 43, 112 (2013)
\bibitem{matsumoto05} T. Matsumoto et al, {\it Infrared Telescope in Space Observations of the Near-Infrared Extragalactic Background Light}, {\apj} 626, 31 (2005)
\bibitem{bernstein07} R.A. Bernstein et al, {\it The First Detections of the Extragalactic Background Light at 3000, 5500, and 8000 {\AA}. III. Cosmological Implications}, {\apj} 571, 107 (2002)
\bibitem{madau00} P. Madau and L. Pozzetti, {\it Deep galaxy counts, extragalactic background light and the stellar baryon budget}, {\mnras} 312, L9 (2000)
\bibitem{dominguez11a} A. Dom\'inguez et al, {\it Extragalactic background light inferred from AEGIS galaxy-SED-type fractions}, {\mnras} 410, 2556 (2011)
\bibitem{aharonian06} F. Aharonian et al, {\it A low level of extragalactic background light as revealed by {$\gamma$}-rays from blazars}, {\nat} 440, 1018 (2006)
\bibitem{ackermann12} M. Ackermann et al, {\it The imprint of the extragalactic background light in the gamma-ray spectra of blazars}, {Science} 338, 1190 (2012)
\bibitem{abramowski13} A. Abramowski et al, {\it Measurement of the extragalactic background light imprint on the spectra of the brightest blazars observed with H.E.S.S.}, {\aap} 550, A4 (2013)
\bibitem{dominguez13a} A. {Dom{\'{\i}}nguez} et al, {\it Detection of the Cosmic {$\gamma$}-Ray Horizon from Multiwavelength Observations of Blazars}, {\apj} 770, 77 (2013)
\bibitem{biteau15} J. Biteau and D.A. Williams {\it The Extragalactic Background Light, the Hubble Constant, and Anomalies: Conclusions from 20 Years of TeV Gamma-ray Observations}, {\apj} 812, 60 (2015)
\bibitem{1es1011magic} M. Ahnen et al, {\it MAGIC observations of the February 2014 flare of 1ES 1011+496 and ensuing constraint of the EBL density}, {\aap} 590, A24 (2016)
\bibitem{matsuoka11} Y. Matsuoka, {\it Cosmic Optical Background: The View from Pioneer 10/11}, {\apj} 736 (2011)
\bibitem{matsuura17} S. Matsuura et al, {\it New Spectral Evidence of an Unaccounted Component of the Near-infrared Extragalactic Background Light from the CIBER}, {\apj} 839, 7 (2017)
\bibitem{mattila17} K. Mattila, {\it Extragalactic background Light: a measurement at 400 nm using dark cloud shadow II. Spectroscopic separation of dark cloud's light, and results}, arXiv:1705.10790 (2017)

\bibitem{magicupgrade1} J. Aleksi\'c et al, {\it The major upgrade of the MAGIC telescopes, Part I: The
hardware improvements and the commissioning of the system} Astroparticle Physics 72, 61 (20a16)
\bibitem{magicupgrade2} J. Aleksi\'c et al, {\it The major upgrade of the MAGIC telescopes, Part II: A
performance study using observations of the Crab Nebula}, Astroparticle Physics, 72, 76 (2016).
\bibitem{pks1441magic} M. Ahnen et al, {\it Very High Energy {$\gamma$}-Rays from the Universe's Middle Age: Detection of the z = 0.940 Blazar PKS 1441+25 with MAGIC}, {\apjl} 815, L23 (2015)
\bibitem{pks1441veritas} A.U. Abeysekara et al, {\it Gamma-Rays from the Quasar PKS 1441+25: Story of an Escape}, {\apjl} 815, L22 (2015)
\bibitem{b0218} M. Ahnen et al, {\it Detection of very high energy gamma-ray emission from the gravitationally lensed blazar QSO B0218+357 with the MAGIC telescopes}, {\aap} 595, A98 (2016)
\bibitem{fermianalysis} https://fermi.gsfc.nasa.gov/ssc/data/analysis/
\bibitem{enrico} D. A. S\'anchez, C. Deil, {\it Enrico : a Python package to simplify Fermi-LAT analysis}. Proc. 33$^{rd}$ ICRC, R\'io de Janeiro, Brazil (2013) arXiv1307.4534
\bibitem{rolke} W. Rolke et al, {\it Limits and confidence intervals in the presence of nuisance parameters}, Nucl.Instrum.Meth.A 551, 493 (2005)
\bibitem{cowan} G. Cowan, {\it Statistical data analysis}, Clarendon Press, Oxford (1998)
\bibitem{minuit2} M. Hatlo et al, Developments of Mathematical Software Libraries for the LHC experiments, IEEE Trans. Nucl. Sci. 52 (2005)
\bibitem{root} R. Brun, F. Rademakers,  ROOT - An Object Oriented Data Analysis Framework, NIM A 389, 81 (1997), http://root.cern.ch/
\bibitem{franceschini08} A. Franceschini et al, {\it Extragalactic optical-infrared background radiation, its time evolution and the cosmic photon-photon opacity}, {\aap} 487, 837 (2008)
\bibitem{gilmore12} R.C. Gilmore et al, {\it Semi-analytic modelling of the extragalactic background light and consequences for extragalactic gamma-ray spectra}, {\mnras} 422, 3189 (2012)
\bibitem{magicgamma2016} D. Mazin et al, {\it EBL Constraints Using a Sample of TeV Gamma-Ray Emitters Measured with the MAGIC Telescopes}, arXiv:1610.09633 (2016)
\bibitem{horns12} D. Horns and M. Meyer, {\it Indications for a pair-production anomaly from the propagation of VHE gamma-rays}, JCAP 2, 33 (2012)
\end{footnotesize}
\vspace{-0.5cm}
\end{thebibliography}
\end{document}